\documentclass[preprint,12pt]{elsarticle}
\usepackage{amssymb}
\usepackage{txfonts}

\newcounter{bla}

\journal{Computer Physics Communications}

\begin{document}

\begin{frontmatter}

\title{$i$QIST v0.7: An open source continuous-time quantum Monte Carlo impurity solver toolkit}

\author[a]{Li Huang\corref{author}}

\cortext[author] {Corresponding author.\\\textit{E-mail address:} lihuang.dmft@gmail.com}
\address[a]{Science and Technology on Surface Physics and Chemistry Laboratory, P.O. Box 9-35, Jiangyou 621908, China}

\begin{abstract}
In this paper, we present a new version of the $i$QIST software package, which is capable of solving various quantum impurity models by using the hybridization expansion (or strong coupling expansion) continuous-time quantum Monte Carlo algorithm. In the revised version, the software architecture is completely redesigned. New basis (intermediate representation or singular value decomposition representation) for the single-particle and two-particle Green's functions is introduced. A lot of useful physical observables are added, such as the charge susceptibility, fidelity susceptibility, Binder cumulant, and autocorrelation time. Especially, we optimize measurement for the two-particle Green's functions. Both the particle-hole and particle-particle channels are supported. In addition, the block structure of the two-particle Green's functions is exploited to accelerate the calculation. Finally, we fix some known bugs and limitations. The computational efficiency of the code is greatly enhanced.     
\end{abstract}


\end{frontmatter}

{\bf \noindent New version program summary}

\begin{small}
\noindent
{\em Program Title:} $i$QIST \\
{\em Licensing provisions:}  GPLv3 \\
{\em Programming language:} Fortran 90, Python, Bash shell \\
{\em Supplementary material:} \\
{\em Journal reference of previous version:} Computer Physics Communications 195, 140 (2015) \\
{\em Does the new version supersede the previous version?:} Yes \\
{\em Reasons for the new version:} Since the release of $i$QIST v0.5 in 2015 [1], we keep improving this code. The new version incorporates many new features, optimizations, and bug-fixes, which will be detailed below. \\
{\em Summary of revisions:} 
\begin{itemize}
\item In the revised version, only the \texttt{narcissus} component and the \texttt{hibiscus} component are retained. The former implements a hybridization expansion continuous-time quantum Monte Carlo impurity solver for density-density type interaction. The latter includes many auxiliary tools and useful scripts. The two components are carefully benchmarked and verified. 
\item New representation (intermediate representation or singular value decomposition representation) for the single-particle and two-particle Green's functions are implemented [2]. It is better than the Legendre orthogonal polynomial representation.
\item Now many physical observables can be measured, including the spin susceptibility, charge susceptibility, fidelity susceptibility, kinetic energy fluctuation, kurtosis and skewness of perturbation expansion order, and Binder cumulant [3]. The fidelity susceptibility can be used to detect the quantum phase transition efficiently.
\item The measurement of two-particle Green's functions is reimplemented. Both the particle-hole and particle-particle channels are supported on the same footing. The block structure and symmetry of the two-particle Green's functions are also utilized to reduce the CPU burden and memory requirement. The $i$QIST software package usually act as a computational engine (i.e, quantum impurity solver) in the calculations of the dynamical mean-field theory and its diagrammatic extensions, such as the dual fermions, dual bosons, and dynamical vertex approximation [4]. The essential inputs for these diagrammatic extensions are the two-particle Green's functions. We design the data structure and file format for the two-particle Green's functions carefully, so that they can be easily accessed by the open source dual fermions code \texttt{opendf} [5].
\item Now the code can output the standard deviations (error bars) for all of the physical observables.  
\item Now the code can measure the autocorrelation function/time for the total occupation number, and then use it to adjust automatically the interval between two successive measurements.  
\item Usually we have to perform analytical continuations for the single-particle Green's function, spin susceptibility, and self-energy function, and convert them from imaginary time or Matsubara frequency space to real-frequency space. These tasks are extremely non-trivial. Some scripts are provided to deal with the output data. They are converted into column-wise files, so that some external codes, such as SpM [6] and $\Omega$MaxEnt [7], can be used to do the analytical continuations.
\item We also provide some scripts to generate initial hybridization function and retarded interaction function, and make animation movie for the random walk in the Monte Carlo configuration space.
\item In the previous version of $i$QIST [1], once the Coulomb interaction $U$ is dynamic and the improved estimator for the self-energy function [8] is employed, the real-part of Matsubara self-energy function is wrong. In the revised version, we fix this severe bug. 
\item We maintain a comprehensive online manual for the code. The users can read the newest $i$QIST's documentation in the following website:

www.gitbook.com/book/huangli712/iqist/.

\item Now the code repository is transferred to Github. The uses can download the newest version of $i$QIST from the following website: 

github.com/huangli712/iqist.

\end{itemize}
{\em Nature of problem(approx. 50-250 words):} In the dynamical mean-field theory and its diagrammatic extensions, the bottleneck is to solve a quantum impurity model self-consistently [4,9]. The quantum impurity model is a Hamiltonian that is used to describe quantum impurities embedded in bath environment or metallic hosts. The goal of the $i$QIST software package is to provide highly effective quantum impurity solvers. \\
{\em Solution method(approx. 50-250 words):} In the $i$QIST software package, we only implement the hybridization expansion continuous-time quantum Monte Carlo algorithm [10]. \\
{\em Additional comments including Restrictions and Unusual features (approx. 50-250 words):} In the revised version, the \texttt{manjushaka} component which implements a hybridization expansion continuous-time quantum Monte Carlo impurity solver for general Coulomb interaction is removed temporally due to numerical unstable problem. The application programming interfaces for Python and Fortran languages are also disabled. They will be released in the next version. \\
{\bf \noindent Acknowledges:} This work was supported by the Natural Science Foundation of China (No.~11504340), Foundation of President of China Academy of Engineering Physics (No.~YZ2015012), and Science Challenge Project of China (No. TZ201604).\\

\end{small}


\begin{thebibliography}{0}
\bibitem{1} Li Huang, Yilin Wang, Zi Yang Meng, Liang Du, Philipp Werner, and Xi Dai, $i$QIST: An open source continuous-time quantum Monte Carlo impurity solver toolkit, Computer Physics Communications 195 (2015), 140.
\bibitem{2} Hiroshi Shinaoka, Junya Otsuki, Masayuki Ohzeki, and Kazuyoshi Yoshimi, Compressing Green's function using intermediate representation between imaginary-time and real-frequency domains, Phys. Rev. B 96 (2017), 035147. 
\bibitem{3} Li Huang, Yilin Wang, Lei Wang and Philipp Werner, Detecting phase transitions and crossovers in Hubbard models using the fidelity susceptibility, Phys. Rev. B 94 (2016), 235110.
\bibitem{4} G. Rohringer, H. Hafermann, A. Toschi, A. A. Katanin, A. E. Antipov, M. I. Katsnelson, A. I. Lichtenstein, A. N. Rubtsov, and K. Held, Diagrammatic routes to non-local correlations beyond dynamical mean field theory, arXiv:1705.00024.
\bibitem{5} Andrey E. Antipov, James P.F. LeBlanc, and Emanuel Gull, opendf - an implementation of the dual fermion method for strongly correlated systems, Phys. Procedia 68 (2015), 43.
\bibitem{6} Junya Otsuki, Masayuki Ohzeki, Hiroshi Shinaoka, and Kazuyoshi Yoshimi, Sparse modeling approach to analytical continuation of imaginary-time quantum Monte Carlo data, Phys. Rev. E 95 (2017), 061302.
\bibitem{7} Dominic Bergeron and A.-M. S. Tremblay, Algorithms for optimized maximum entropy and diagnostic tools for analytic continuation, Phys. Rev. E 94 (2016), 023303.
\bibitem{8} Hartmut Hafermann, Self-energy and vertex functions from hybridization-expansion continuous-time quantum Monte Carlo for impurity models with retarded interaction, Phys. Rev. B 89 (2014), 235128.
\bibitem{9} Antoine Georges, Gabriel Kotliar, Werner Krauth, and Marcelo J. Rozenberg, Dynamical mean-field theory of strongly correlated fermion systems and the limit of infinite dimensions, Rev. Mod. Phys. 68 (1996), 13.
\bibitem{10} Emanuel Gull, Andrew J. Millis, Alexander I. Lichtenstein, Alexey N. Rubtsov, Matthias Troyer, and Philipp Werner, Continuous-time Monte Carlo methods for quantum impurity models, Rev. Mod. Phys. 83 (2011), 349.
\end{thebibliography}
\end{document}